\documentclass[12pt]{article}

\newcommand{\add}{\addtocounter{eqncnt}{1}}
\newcounter{eqncnt}[section]

\newcommand{\be}{\begin{equation}}
\newcommand{\ee}{\end{equation}\add}
\newcommand{\bea}{\begin{eqnarray}}
\newcommand{\eea}{\end{eqnarray}}

\renewcommand{\d}{\mathrm{d}}

\begin{document}
\begin{center}
{\Large \bf Note on the invariant classification of vacuum type D spacetimes } \\[2mm]

\vskip .5in

{\sc A. Coley}\\
{\it Department of Mathematics and Statistics}\\
{\it Dalhousie University, Halifax, NS B3H 3J5, Canada}\\
{\it aac@mathstat.dal.ca }

\vskip .5in
{\sc and}\\

\vskip .5in
{\sc S. Hervik}\\
{\it Faculty of Science and Technology}\\
{\it University of Stavanger, N-4036 Stavanger, Norway}\\
{\it sigbjorn.hervik@uis.no}\\

\end{center}

\begin{abstract}

We illustrate the fact that the class of 
vacuum type D spacetimes which are $\mathcal{I}$-\emph{non-degenerate}
are invariantly classified by their scalar polynomial  curvature invariants. 

\end{abstract}


\newpage

\section{Introduction}
In Lorentzian spacetimes, identical
metrics  are
often given in different coordinate systems, which
disguises their true equivalence. It is consequently of fundamental
importance to have an invariant way to distinguish spacetime
metrics. The invariant classification
developed by Karlhede \cite{Karlhede}, based 
on the Newman-Penrose (NP) formalism  \cite{NP} 
and the Cartan equivalence method, is widely used to
characterize Lorentzian spacetimes in terms of their Cartan scalars in general relativity
\cite{kramer}.

However, perhaps the easiest way of distinguishing metrics is
through their {\em scalar polynomial curvature invariants}, which are scalars obtained
by contraction from a polynomial in the Riemann tensor and its
covariant derivatives, due to the
fact that inequivalent invariants implies inequivalent metrics. 
In \cite{inv} the class of four-dimensional
Lorentzian manifolds that can be completely characterized locally (in the sense defined below) by their
scalar polynomial curvature invariants was determined.
In particular, for any given Lorentzian spacetime,
$(\mathcal{M},g)$, let  $\mathcal{I}$ denote the set of all
scalar polynomial curvature invariants constructed from the Riemann tensor
and its covariant derivatives. If there does not exist a
metric deformation of $g$ having the same set of invariants as
$g$, then we call the set of invariants $\mathcal{I}$-\emph{non-degenerate},
and the spacetime metric $g$ is called
\emph{$\mathcal{I}$-non-degenerate}. This means that for a metric
which is $\mathcal{I}$-non-degenerate, the invariants locally
characterize the spacetime uniquely. In \cite{inv} it was proven
that a four-dimensional Lorentzian spacetime metric is either
\emph{$\mathcal{I}$-non-degenerate} or \emph{degenerate Kundt}.

This important result implies that metrics not determined by their
scalar polynomial curvature invariants  (at least locally) must be
of degenerate Kundt form. These Kundt metrics therefore correspond to degenerate
metrics in the sense that many such spacetimes can have identical
scalar polynomial invariants.  The Kundt class is defined as those spacetimes admitting a
null vector $\ell$ that is geodesic, expansion-free, shear-free and twist-free \cite{kramer,Kundt}. 
It follows that there
exists a {\em kinematic} frame in which the NP scalars
$\kappa=\sigma=\rho=\epsilon$ all vanish.
In a {\it degenerate Kundt} spacetime there exists
a common null frame in which the geodesic, expansion-free,
shear-free and twist-free null vector $\ell$ is also the null
vector in which all positive boost weight terms of the Riemann
tensor and its covariant derivatives are zero  \cite{Kundt}. 
Any metric in the degenerate Kundt class can be written in a canonical form 
\cite{Kundt,class}.

Clearly, by knowing which spacetimes can be characterized
by their scalar curvature invariants alone, the computations
of the invariants (i.e., simple polynomial scalar invariants) is much more
straightforward and can be done algorithmically (i.e., the full
complexity of the equivalence method is not necessary). On the other
hand, the Cartan equivalence method also contains, at least in principle,
the conditions under which the classification is complete (although
in practice carrying out the classification for the more general
spacetimes can be very difficult). Therefore, in a sense,
the full machinery of the Cartan equivalence method is only necessary
for the classification of the degenerate Kundt spacetimes \cite{milson}.

\section{Vacuum type D spacetimes}

In this note we shall illustrate this by considering the class of 
vacuum type D spacetimes.  These spacetimes were studied by Kinnersley
\cite{Kinnersley}, who divided the spacetimes into 4 subclasses (I-IV)
(which subdivides further into 10 distinct classes of metrics). This simple class
of spacetimes are of physical interest, and have been classified using the
Karlhede method \cite{Aman}. Recently, an earlier invariant classification of the
vacuum type D spacetimes based  on the NP formalism \cite{Collins} has been extended 
by exploiting some NP identities using the GHP formalism \cite{Edgar}. Clearly the classification of 
vacuum type D spacetimes is still of interest.
We shall show that we can classify these spacetimes, which are, in general, 
$\mathcal{I}$-\emph{non-degenerate} \cite{inv}, using
simple scalar polynomial curvature invariants. As an illustration we shall
present the results for the Kinnersley class I spacetimes
(the other cases work in a similar way). Note that this method does not
reproduce the determination of the upper bound (2) on the order of the Cartan
scalars for these spacetimes \cite{Edgar,Aman}. 

\subsection{Kinnersley class I metric}
The Kinnersley class I Petrov type D vacuum metric is given by \cite{Kinnersley}:
\begin{eqnarray} \d s^2 &=&  -2S\d u^2  +2\d u\d r +4D^{-1}
lSy\d u\d x -4D^{-1}  lSx\d u\d y -2D^{-1}  ly\d r\d x +2D^{-1}  lx\d r\d y\nonumber \\
&& +\left(-2D^{-2}  l^2 Sy^2 -D^{-2}  zz^* \right)\d x^2  +4D^{-2}  
l^2 Sxy\d x\d y +\left(-2D^{-2}  l^2 Sx^2 -D^{-2}  zz^* \right)\d y^2, \nonumber 
\end{eqnarray}
where the variables (labelled $0-3$) are $(u,r,x,y)$, $z = r + il$ is a complex variable,
$l,~ C, ~(2Cil + m)$ and $(-2Cil + m)$ are constants, and 
\begin{eqnarray}
S& \equiv&  2Cl^2{(z z^{*})}^{-1} - C + \frac{1}{2}(2Cil + m) r {(z z^{*})}^{-1}
+ \frac{1}{2}(-2Cil + m) r {(z z^{*})}^{-1}\nonumber \\
D& \equiv& \frac{1}{2}C x^2 + \frac{1}{2}C y^2 + 1. \nonumber
\end{eqnarray}

There are 4 algebraically independent (complex) Cartan invariants, which are 
{\footnote{ The worksheets, based on 
the work of J. {\AA}man, were kindly provided by J. Skea. We follow the notation
of these worksheets and \cite{Aman}. Note that
$Re({\Psi}_2) = (2Cil+m)z^{-3} + (-2Cil+m){z^{*}}^{-3} $ is an independent function.}}:
\begin{eqnarray}       
{\Psi}_2  &=&  -(2Cil+m)z^{-3}\nonumber \\
{\nabla}{\Psi}_{20'} &=& 3(2Cil+m)S^{\frac{1}{2}} z^{-4} = {\nabla}{\Psi}_{31'} \nonumber \\
{\nabla}^2 {\Psi}_{20'} &=& -12(2Cil+m)Sz^{-5} = {\nabla}^2 {\Psi}_{42'}\nonumber \\                       
{\nabla}^2 {\Psi}_{31'}    &=&  -3C(2Cil+m)z^{-4}  {z^*}^{-1}   
+\frac{3}{2}(2Cil+m)^2 z^{-5}  z^{*-1}    -3(2Cil+m)Sz^{-4}  z^{*-1}\nonumber \\
&& -12(2Cil+m)Sz^{-5}.\nonumber 
\end{eqnarray}

We calculate the 4 (complex) scalar polynomial invariants:
\begin{eqnarray}
I &\equiv& \frac 12\Psi_{abcd}\Psi^{abcd}=3(2Cil+m)^2 z^{-6}, \nonumber \\      
C^{\alpha\beta\gamma\delta}C_{\alpha\beta\gamma\delta}&=& 24(2Cil+m)^2 z^{-6} +24(-2Cil+m)^2 z^{*-6}\nonumber \\
{\Psi}^{(abcd;e)f'} {\Psi}_{(abcd;e)f'}  &=& 180(2Cil+m)^2 Sz^{-8}\nonumber \\                        
C^{\alpha\beta\gamma\delta;\mu}C_{\alpha\beta\gamma\delta;\mu}&=& 720(2Cil+m)^2 Sz^{-8} +720(-2Cil+m)^2 Sz^{*-8}\nonumber, 
\end{eqnarray}
where $\Psi_{abcd}$ is the Weyl spinor, $a,b,..$ are spinor indices, and $\alpha, \beta,...$ are frame indices. 
        
We see that $({\Psi}_2)^2 \sim I$, $({\nabla}{\Psi}_{20'})^2 \sim
{\Psi}^{(abcd;e)f'} {\Psi}_{(abcd;e)f'} $, and $\nabla^2\Psi_{20'}\sim {\Psi}^{(abcd;e)f'} {\Psi}_{(abcd;e)f'}/I$. These can be used to solve for the 4 real parameters $(r,l,C,m)$ in terms of the scalar polynomial invariants. Then by inserting these into the expression for ${\nabla}^2 {\Psi}_{31'}$ 
we can write also this Cartan invariant in terms of polynomial invariants.
Therefore, 
all of the Cartan invariants can be expressed in terms of scalar polynomials.

\section{Discussion}

All of the other cases in the Kinnersley classes can be dealt with in a similar way
\cite{Kinnersley}. The Schwarzschild vacuum type $D$ spacetime belongs to the
Kinnersley class I and, as discussed in \cite{inv}, in canonical
coordinates there are two functionally independent (as functions of the two parameters $r$ and $M$) scalar polynomial
invariants \footnote{ By setting $l=0$, $C=-(1/2)(-m/M)^{2/3}$, 
and by rescaling the radial coordinate, $r\mapsto (-m/M)^{1/3}r$, 
we see that the invariants in the Schwarzschild metric
are equivalent to those in the Kinnersley metric (where $l=0$ is 
the Schwarzschild case).}, 
$C^{\alpha\beta\gamma\delta}C_{\alpha\beta\gamma\delta} = 48{M^2}{r^{-6}}$ and
$C^{\alpha\beta\gamma\delta;\mu}C_{\alpha\beta\gamma\delta;\mu} = 720(r-2M){M^2}{r^{-9}}$, 
and all of the
algebraically independent Cartan scalars
$\Psi_{2}$, $\nabla^2 \Psi_{20'}$, $\nabla^2 \Psi_{31'}$, and
$\nabla^2 \Psi_{42'}$  are related to these two polynomial
curvature invariants (see also 
\cite{Ferr2} for the invariant classification of the Schwarzschild spacetime). 
The Kerr solution belongs to Kinnersley class IIA; this spacetime has been 
invariantly characterized intrinsically (using, in addition to
algebraic scalar invariants, differential Weyl concomitants) \cite{Ferr1}.

In general,  the vacuum type $D$ spacetimes in the Kinnersley
classes are $\mathcal{I}$-\emph{non-degenerate}; from the Bianchi identities
the covariant derivative of the Weyl tensor is of algebraic type I
(for non-flat
spacetimes) except in some very special cases (i.e., some special degenerate Kundt spacetimes
are included in the Kinnersley
classes \cite{Kinnersley}). There is a sense in which the type $D^k$ degenerate
Kundt spacetimes can also be completely characterized by their
scalar polynomial curvature invariants  \cite{inv,Kundt}; however, the
vacuum type $D^k$ spacetimes are trivial.


\paragraph{Acknowledgements:} 

We would like to thank J. Skea for helpful comments.
This work was supported in part by NSERC of Canada.

\end{document}